\documentclass[a4paper,11pt]{article}
\pdfoutput=1 

\usepackage{jheppub} 

\title{\boldmath Systematic uncertainties in integrated luminosity measurement at CEPC}

\author[a,1]{I. Smiljanic,\note{Corresponding author.}}
\author[a]{I. Bozovic Jelisavcic,}
\author[a]{G. Kacarevic,}
\author[a]{N. Vukasinovic,}
\author[a]{G. Milutinovic Dumbelovic,}
\author[a]{T. Agatonovic Jovin,}
\author[a]{I. Vidakovic,}
\author[a]{V. Rekovic,}
\author[b]{J. Stevanovic}
\author[b]{and M. Radulovic}

\affiliation[a]{Vinca Institute of Nuclear Sciences, National Institute of the Republic of Serbia, University of Belgrade\\M. Petrovica Alasa 12-14, Belgrade, Serbia}
\affiliation[b]{University of Kragujevac, Faculty of Science, Department of Physics\\Radoja Domanovica 12, Kragujevac, Serbia}

\emailAdd{i.smiljanic@vinca.rs}
\emailAdd{ibozovic@vinca.rs}
\emailAdd{kacarevicgoran@vinca.rs}
\emailAdd{nvukasinovic@vinca.rs}
\emailAdd{gordanamd@vinca.rs}
\emailAdd{tatjana.jovin@vinca.rs}
\emailAdd{ivana.vidakovic@vinca.rs}
\emailAdd{vladimir.rekovic@cern.ch}
\emailAdd{jasnas@kg.ac.rs}
\emailAdd{mirko.radulovic@pmf.kg.ac.rs}

\abstract{The very forward region of a detector is one of the most challenging regions to instrument at a future $e^{+}e^{-}$ collider. At CEPC, machine-detector interface includes, among others, a calorimeter dedicated for precision measurement of the integrated luminosity at a permille level or better. Here we review a feasibility of such precision, from the point of view of systematic effects arising from the luminometer's mechanical precision and positioning, beam-related requirements and physics background. A method of the beam energy spread determination, initially proposed for FCC, is also discussed for the CEPC beams.}


\begin{document} 
\maketitle
\flushbottom

\section{Introduction}
\label{sec:intro}

The Circular Electron Positron Collider (CEPC) is a large international scientific facility proposed by the Chinese particle physics community in 2012 to test the validity scale of the Standard Model (SM) through precision measurements in the Higgs, BSM and EW sectors. These measurements should provide critical tests of the underlying fundamental physics principles of the Standard Model and are vital in exploration of New Physics beyond the SM. In electron-positron collisions, the CEPC is designed to operate at 91.2 GeV as a $Z^{0}$ factory, at 160 GeV (WW production threshold) and at 240 GeV as a Higgs factory. The vast amount of bottom quarks, charm quarks and $\tau$-leptons produced in $Z^{0}$ decays also makes the CEPC an effective b-factory and $\tau$ and charm factory \cite{CEPC_CDR}.\\
 
In order to achieve precision required for realization of the CEPC physics program, relative uncertainty of the integrated luminosity measurement should be of order of $10^{-4}$ at 91.2 GeV and of order of $10^{-3}$ at 240 GeV. Precision reconstruction of position and energy of electromagnetic showers generated by Bhabha scattering at a high-energy $e^{+}e^{-}$ collider can be achieved with finely granulated compact luminometer. The method for integrated luminosity measurement at CEPC is described in \cite{CEPC_CDR}. However, the precision reconstruction of Bhabha scattering doesn't exhaust the long list of systematic uncertainties of the integrated luminosity measurement, which includes detector related uncertainties, beam related uncertainties and uncertainties originating from physics and machine-related interactions. The latest, that includes beam-beam interactions and beam-gas scattering, is not discussed here. In this paper we review the effects of the detector and beam related uncertainties, namely mechanical uncertainties of the luminometer positioning and size and uncertainties related to the beam energy, beam synchronization and interaction point (IP) displacements, as well as the uncertainty that originates from the miscount of physics background. In addition, motivated by \cite{Janot}, we discuss the possibility of CEPC beam energy spread determination employing the processes in the central tracker.\\

\section{Forward region of CEPC}
\label{sec:forward_region}

\begin{figure}[tbp]
	\centering 
	\includegraphics[width=.49\textwidth]{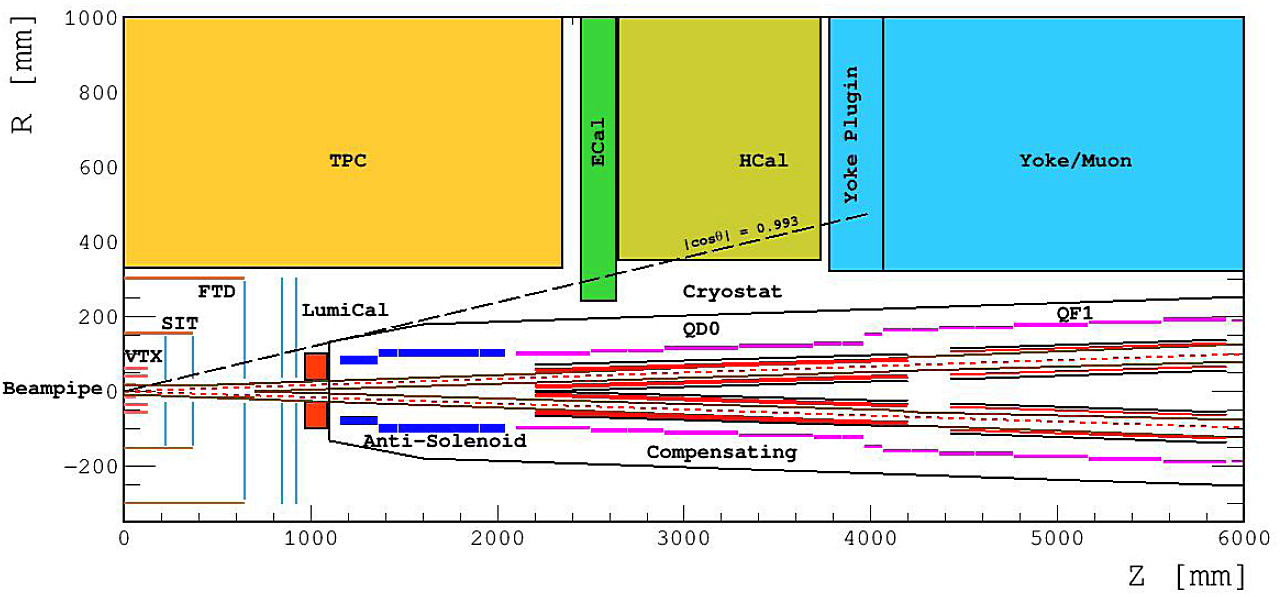}
	\hfill
	\includegraphics[width=.41\textwidth]{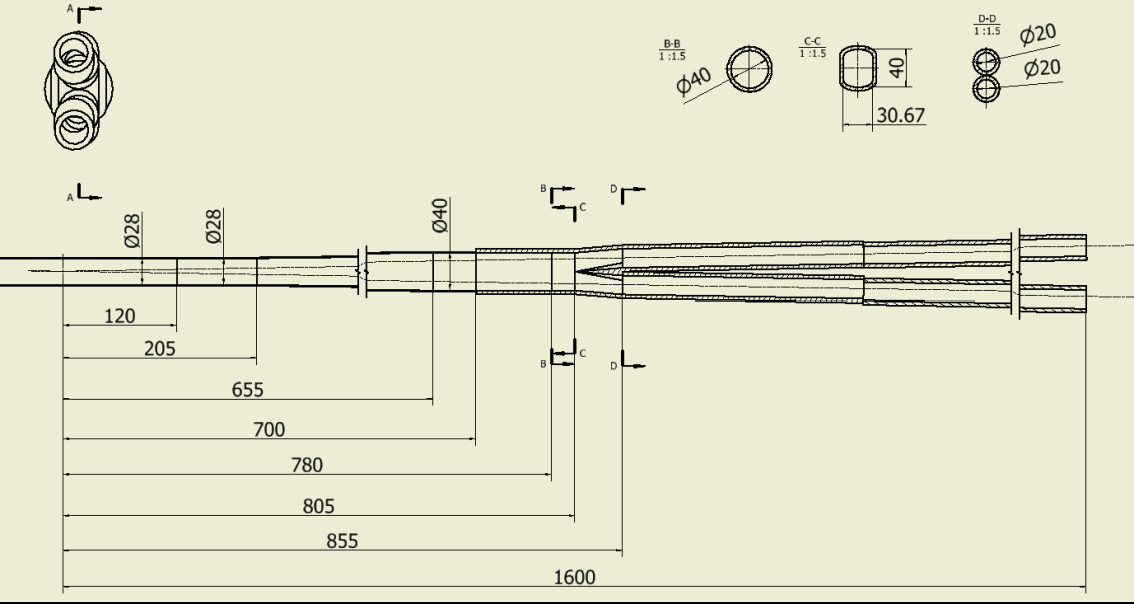}
	\caption{\label{fig:1} Left: Layout of the very forward region at CEPC; Right: Revised beam-pipe design.}
\end{figure}

The Machine Detector Interface (MDI) of CEPC \cite{CEPC_CDR} is about 6 m far from the IP. The accelerator components inside the detector without shielding are within a conical space with an opening angle of arccos 0.993. The two beams collide at the IP with a crossing angle of 33 mrad in the horizontal plane, and the final focus length is 2.2 m. Luminometer at CEPC is proposed to cover the polar angle region between 26 mrad and 105 mrad (with fiducial volume between 53 mrad and 79 mrad) corresponding to the detector aperture of 28.5 mm for the inner radius and 100 mm for the outer, at 950 mm distance from the IP. Luminometer might be supplemented with an additional front layer of tracker in order to improve $e-\gamma$ separation and calibration of the device. Since the luminometer will be placed at $z = \pm 950$ mm, shower leakage from the outer edge of the luminometer, possibly contaminating tracking detectors, have been studied and proven to be negligible after absorption by a 5 mm iron filter positioned around the luminometer \cite{CEPC_CDR}. Layout of the very forward region at CEPC and the revised beam-pipe design are given in Figure \ref{fig:1} \cite{CEPC_CDR}.\\

\begin{figure}[tbp]
	\centering 
	\includegraphics[width=.3\textwidth]{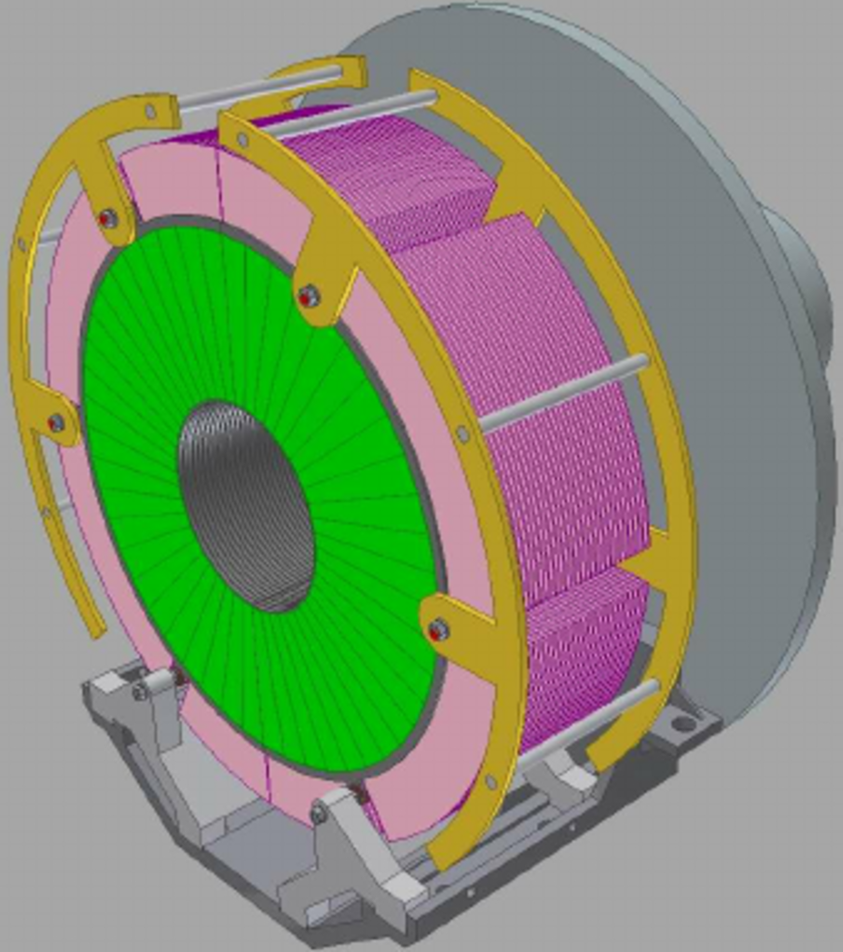}
	\hfill
	\caption{\label{fig:2} Model of Si-W sandwich type of calorimeter developed by FCAL Collaboration, that might be employed at CEPC.}
\end{figure}

It is clear that luminometer must be a compact calorimeter providing shower containment in a longitudinal dimension not larger than 10 cm. Luminometer technology options are still open. Among the candidates for the luminometer are CMS-like shashlik type of calorimeter based on Lutetium Yttrium Orthosilicate ($Lu_{x}Y_{2-x}SiO_{5}:Ce$ - LYSO) \cite{LYSO} and SciFi spaghetti calorimeter with individually read-out fibers (prototyped for J-PARC KL experiment \cite{JParc}). However, the most compact design currently proposed seems to be Si-W sandwich type of calorimeter that could provide over 20 $X_{0}$ in the foreseen longitudinal dimension (Figure \ref{fig:2} \cite{lumical}). The prototype of such a calorimeter has been developed by FCAL Collaboration and has been tested in several test beam campaigns over the last 10 years \cite{Moliere}. This luminometer's fiducial volume ranges from 53 mrad to 79 mrad if used at CEPC, with sensors placed in 2 mm air gaps and it should have fine Si-pixel segmentation (i.e 48/64 azimuthal/radial) and small (effective) Moliere radius ($\sim$2 cm), which would result in excellent resolution in energy and polar angle of the reconstructed electrons provided by the sampling term $\alpha$=0.2 and $\sigma_{\theta} \sim 10^{-2}$ mrad \cite{lumical}. \\

\section{Integrated luminosity measurement and systematic uncertainties}
\label{sec:IntLumiMeasure}

Integrated luminosity measurement is a counting experiment based on Bhabha scattering. It is defined as $L=N_{bh}/\sigma$, where $N_{bh}$ is Bhabha count in the considered interval of time and the certain phase space within the detector acceptance (fiducial) region and $\sigma$ is the theoretical cross-section in the same geometrical and phase space. However, in a real experiment there are several effects influencing Bhabha count. Here we list some of them that are addressed at the simulation level, assuming detector geometry as described in Section \ref{sec:forward_region} and the CEPC beams as in \cite{CEPC_CDR}: uncertainties from mechanics (detector manufacturing, positioning and alignment); uncertainties from the beam properties (center-of-mass energy, beam-energy asymmetry, beam synchronization, IP displacements) and Bhabha miscounts from two-photon processes as the main physics background. In order to control integrated luminosity with the relative precision of $10^{-4}$ ($10^{-3}$), both $N_{bh}$ and $\sigma$ should be known at the same order of magnitude, which means that all these effects should be controlled with the same precision.\\

\subsection{Uncertainties stemming from mechanics and positioning}
\label{subsec:UncMechPos}

Systematic uncertainties from detector and machine-detector interface related effects have been quantified through a simulation study, assuming $10^{7}$ Bhabha scattering events generated using BHLUMI V4.04 Bhabha event generator \cite{BHLUMI}, at two center-of-mass energies: 240 GeV and $Z^{0}$ production threshold. Detector fiducial volume, where the showers are fully contained and thus the sampling term constant, is assumed as described in Section \ref{sec:forward_region}, with the luminometer placed at ongoing beams. The effective Bhabha cross-section in this angular range is of order of a few nb. Final state particles are generated in the polar angle range from 45 mrad to 85 mrad, somewhat broader than the detector fiducial volume, in order to allow events with non-collinear Final State Radiation (FSR) to contribute. Close-by particles are summed up to imitate cluster merging. We assume that the shower leakage from the luminometer is negligible.\\

\begin{figure}[tbp]
	\centering 
	\includegraphics[width=.5\textwidth]{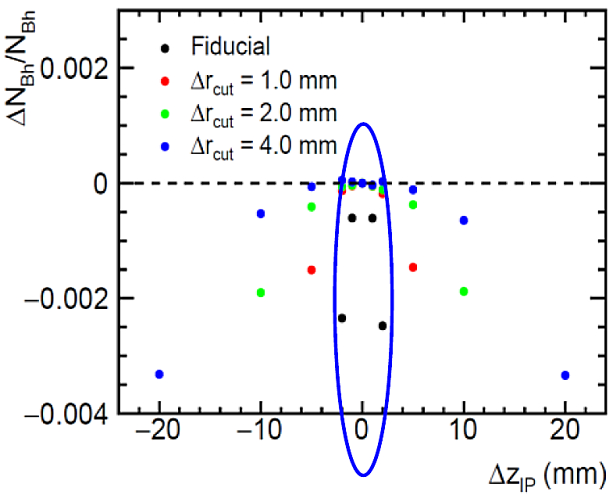}
	\hfill
	\caption{\label{fig:3} Luminosity uncertainty from the longitudinal IP displacements ($\Delta Z_{IP}$) w.r.t. the luminometer, for symmetric (circled) and asymmetric selection with a radial shrink of the fiducial volume $\Delta r$.}
\end{figure}

Furthermore, we have applied event selection that is asymmetric in polar angle acceptance on the left and right arm of the detector, as it has been done at OPAL \cite{OPAL}. That is, at one side we consider the full fiducial volume, while at the other side we shrink the radial acceptance for $\Delta r$. This has been done subsequently to the left (L) and right (R) side of the luminometer, on event by event basis. In addition, we require high-energy electrons carrying above 50\% of the available beam energy. Against this type of event selection for luminosity measurement, we compare the selection based of the full fiducial volumes on both sides of the detector. An example is given in Figure \ref{fig:3}, illustrating the cancelation of systematics uncertainties caused by the assumption of L-R symmetry in an event, when asymmetric selection in polar angle is applied. It is clear that asymmetric selection is advantageous, requiring a luminometer placed at the outgoing beams.\\

Considered detector-related uncertainties arising from manufacturing, positioning and alignment are:

\begin{description}
	\item[$\bullet$]uncertainty of the luminometer inner radius ($\Delta r_{in}$),
	\item[$\bullet$]spread of the measured radial shower position w.r.t. the true impact position on the luminometer front plane ($\sigma_{r}$),
	\item[$\bullet$]uncertainty of the longitudinal distance between left and right halves of the luminometer ($\Delta l$),
	\item[$\bullet$]radial and axial ($\sigma_{x_{IP}}, \sigma_{z_{IP}}$) mechanical fluctuations of the luminometer position with respect to the IP, caused by vibrations and thermal stress,
	\item[$\bullet$]twist of the calorimeters corresponding to different rotations of the left and right detector axis with respect to the outgoing beam ($\Delta \varphi$).
\end{description}

We assumed $10^{-3}$ and $10^{-4}$ contribution to the relative uncertainty of integrated luminosity from each individual effect, at 240 GeV and $Z^{0}$ pole respectively. Table \ref{tab:1} gives corresponding requirements of the listed parameters. It is clear that due to the $\sim 1/\theta^{3}$ dependence of the Bhabha cross-section from the polar angle, inner aperture of the luminometer is one of the most demanding mechanical parameters to control. The system for detector positioning measurement must be such to provide $\sim$80 $\mu$m uncertainty in distance between L and R halves of luminometer.

\begin{table}[tbp]
	\centering
	\caption{\label{tab:1} Required absolute precision of mechanical parameters contributing to the relative uncertainty of the integrated luminosity of $10^{-3}$ ($10^{-4}$) at 240 GeV center-of-mass energy ($Z^{0}$ pole).}
	\small
\begin{tabular}{|l|c|c|}
	\hline 
	\textbf{parameter} & \textbf{precision @ 240 GeV } & \textbf{precision @ 91 GeV } \\ 
	\hline 

	$\Delta r_{in}$ ($\mu$m) & 10 & 1 \\ 
	\hline 
	$\sigma_{r}$ (mm) & 1.00 & 0.20 \\ 
	\hline 
	$\Delta l$ (mm) & 1.00 & 0.08 \\ 
	\hline 
	$\sigma_{x_{IP}}$ (mm) & 1.0 & 0.5 \\ 
	\hline 
	$\sigma_{z_{IP}}$ (mm) & 10 & 7 \\ 
	\hline 
	$\Delta \varphi$ (mrad) & 6.0 & 0.8 \\ 
	\hline 
\end{tabular} 
\end{table}

\subsection{MDI related uncertainties}
\label{subsec:MDIunc}

Several uncertainties are considered that may arise from the beam properties and its delivery to the interaction point:

\begin{description}
	\item[$\bullet$]uncertainty of the average net center-of-mass energy ($\Delta E_{CM}$),
	\item[$\bullet$]asymmetry in energy of the $e^{+}$ and $e^{-}$ beams $\lvert E_{e^{+}}-E_{e^{-}}\rvert$ given as the maximal deviation of the beam energy from its nominal value ($\Delta E$),
	\item[$\bullet$]radial  ($\Delta x_{IP}$) and axial  ($\Delta z_{IP}$) IP position displacements with respect to the luminometer, caused by the finite transverse beam sizes and beam synchronization respectively,
	\item[$\bullet$]time shift in beam synchronization ($\tau$) leading to IP longitudinal displacement  $\Delta z_{IP}$. 
\end{description}

Table \ref{tab:2} gives absolute uncertainties of these parameters contributing to the relative uncertainty of integrated luminosity in the same way as described for Table \ref{tab:1}.\\

\begin{table}[tbp]
	\centering
	\caption{\label{tab:2} {Required absolute precision of MDI parameters contributing to the relative uncertainty of the integrated luminosity of  $10^{-3}$ ($10^{-4}$) at 240 GeV center-of-mass energy ($Z^{0}$ pole).}}
	\small
	\begin{tabular}{|l|c|c|}
		\hline 
		\textbf{parameter} & \textbf{precision @ 240 GeV } & \textbf{precision @ 91 GeV } \\ 
		\hline 
		$\Delta E_{CM}$ (MeV) & 240 & 9 \\ 
		\hline 
		$\Delta E$ (MeV) & 120 & 5 \\ 
		\hline 
		$\Delta x_{IP}$ (mm) & 1.0 & 0.5 \\ 
		\hline 
		$\Delta z_{IP}$ (mm) & 10.0 & 2.0 \\ 
		\hline 
		$\tau$ (ps) & 15 & 3 \\ 
		\hline 
\end{tabular} 
\end{table}

\begin{figure}[tbp]
	\centering 
	\includegraphics[width=.5\textwidth]{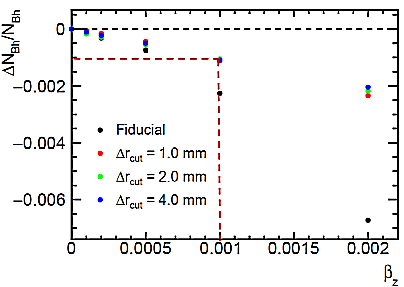}
	\hfill
	\caption{\label{fig:4} Loss of the Bhabha count in the luminometer due to the longitudinal boost of the CM frame $\beta_{Z}$, where $\beta_{Z}$ is given as $2\Delta E/E_{CM}$. $\Delta r$ values correspond to different polar angle acceptance of detector left and right arms. Red dashed line indicates $10^{-3}$ uncertainty in count required at 240 GeV CEPC.}
\end{figure}

Another challenge comes from the uncertainty of energy of individual beams (and of the effective center-of-mass energy) needed to be controlled at the level of $\sim10^{-5}$ with respect to the nominal beam energy at the $Z^{0}$ pole, that is smaller than the foreseen beam energy spread of 0.08\% \cite{CEPC_CDR}. The current value of the beam energy spread at the $Z^{0}$ pole will contribute to the relative uncertainty of the Bhabha count maximally as $1.6\cdot{10^{-3}}$, through the uncertainty of the effective center-of-mass energy $\Delta E_{CM}$ for the Bhabha cross-section calculation ($\sigma_{Bh}\sim1/E_{CM}^{2}$), or as the asymmetry in beam energies giving rise to longitudinal boost of the center-of-mass (collision) system (CM) w.r.t. the laboratory frame. Figure \ref{fig:4} illustrates counting loss in the luminometer due to longitudinal boost of the CM frame $\beta_{Z} \sim 2\Delta E$, at 240 GeV center-of-mass energy.

\subsection{Two-photon processes as a background}
\label{subsec:2photonBcg}

\begin{figure}[tbp]
	\centering 
	\includegraphics[width=.5\textwidth]{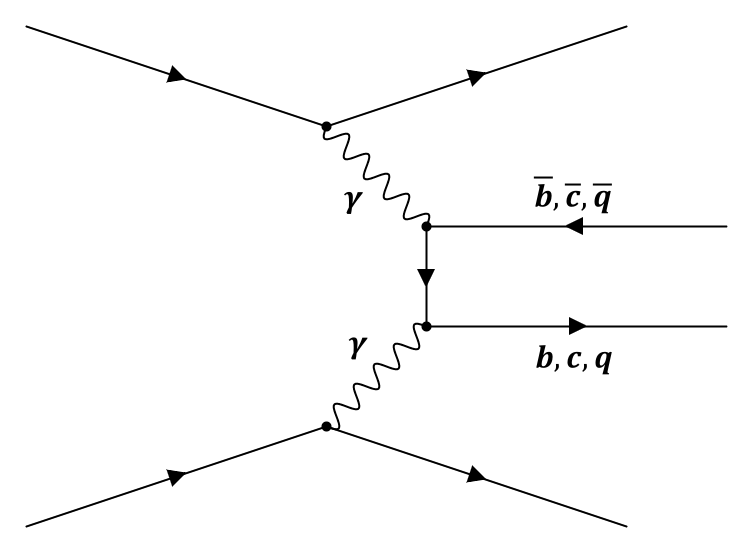}
	\hfill
	\caption{\label{fig:5} Feynman diagram of two-photon process producing $e^{+}e^{-} + 2-jet$ final state in electron-positron collisions.}
\end{figure}

In $e^{+}e^{-}$ collisions there are several 4-fermion proceses (multiperipheral, annihilation, conversion and bremsstrahlung) representing possible background for the Bhabha scattering. The multiperipheral (two-photon) process, given in Figure \ref{fig:5}, should be considered due to its large cross-section ($\sim$nb) and to the fact that spectator electrons are emitted at very small polar angles. Even though the most of high-energy electron spectators from these processes go below the luminometer's angular acceptance region, some of them can still be misidentified as Bhabha electrons. Here we quantify their contribution, assuming geometrical parameters as in Section \ref{sec:forward_region}. \\

\begin{figure}[tbp]
	\centering 
	\includegraphics[width=.50\textwidth]{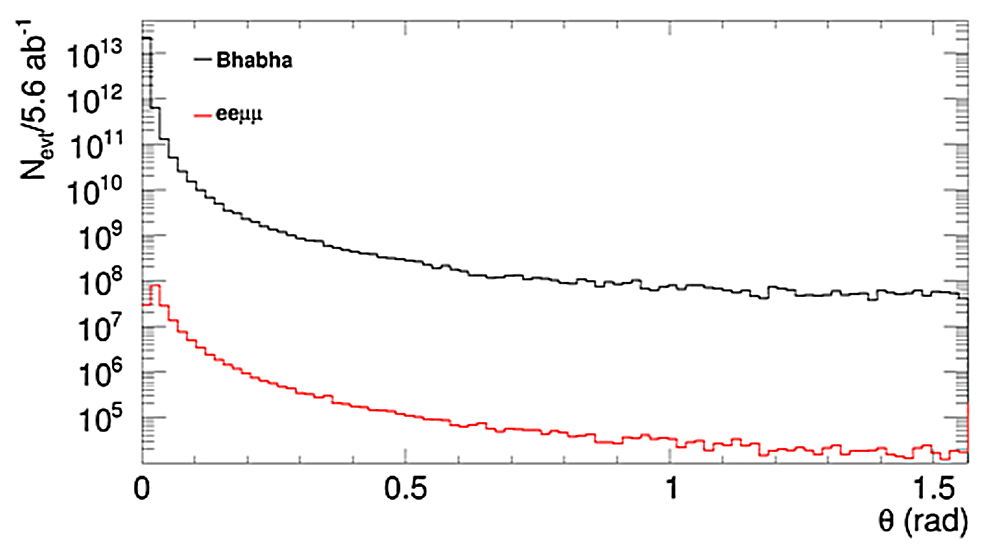}
	\caption{\label{fig:6} Normalized polar angle distribution of signal (black line) and background (red line) events.}
\end{figure}

In order to estimate the background to signal ratio at 240 GeV CEPC (where 2-photon processes are more abundant than at the $Z^{0}$ pole, due to $ln(s)$ cross-section dependence), we simulated $10^{5}$ $e^{+}e^{-} \rightarrow e^{+}e^{-}\mu^{+}\mu^{-}$ events using WHIZARD V2.6 \cite{WHIZARD} in the polar angle range $\lvert \cos \theta \rvert<0.999$. The effective cross-section $\sigma_{eff}\sim0.3$ pb is found in the fiducial volume of the luminometer. $10^{7}$ Bhabha events are simulated using BHLUMI V4.04 in the polar angle range $\theta>3$ mrad, with $\sigma_{eff}\sim3.3$ nb in the fiducial volume of the luminometer. Polar angle distribution for signal and background is given in Figure \ref{fig:6}. Counts are normalized to $5.6$ ab$^{-1}$, which corresponds to 7 years of data taking at 240 GeV CEPC.\\

From Figure \ref{fig:6} it is clear that the most of spectators go below the luminometer, while contamination of the signal in the luminometer's fiducial volume is of order of $10^{-4}$ even without any event selection. In this study we've looked into spectator electrons in the luminometer fiducial volume, from the $e^{+}e^{-} \rightarrow e^{+}e^{-}\mu^{+}\mu^{-}$ final state of 2-photon processes. We found that the total amount of background should be conservatively scaled by a factor 3 to account for flavor integration. Further refinements are possible with the coplanarity request between left and right detector arms ($\lvert\varphi_{e^{+}}-\varphi_{e^{-}}\rvert$), as well as the cut on relative energy $E_{rel}=\lvert E_{e^{+}}+E_{e^{-}}\rvert/2E_{beam}$ that is also useful to suppress both 2-photon background and off-momentum particles. Finally, physics background can be always taken as a correction to the Bhabha count if the uncertainty of its cross-section is available.\\

\section{Determination of the beam energy spread}
\label{sec:spread}

Motivated by the similar work done at FCC-ee \cite{Janot}, and having in mind that precision of several EW observables at the $Z^{0}$ pole (such as the $Z^{0}$ mass, width and production cross-section) depends on the precision of beam energy spread determination, we looked into possibility to measure it at CEPC using well defined central process like di-muon production $e^{+}e^{-} \rightarrow \mu^{+}\mu^{-}$. Having in mind the projected performance of the central tracker to efficiently ($\sim99\%$) and precisely (100 $\mu$m position resolution) reconstruct muons \cite{CEPC_CDR}, process $e^{+}e^{-} \rightarrow \mu^{+}\mu^{-}$ with a cross-section of 1.5 nb at the $Z^{0}$ pole seems to be an optimal choice.\\

We'll argue that the effective center-of-mass energy ($s'$) is sensitive to variation of the beam energy spread that consequently can be determined from the population of the peak of the $s'$ distribution. In order to determine $s'$ sensitivity to the beam energy spread, we generated several hundreds thousands $e^{+}e^{-} \rightarrow \mu^{+}\mu^{-}$ events at 91.2 GeV and 240 GeV center-of-mass energies. Events are generated using WHIZARD 2.6, in the polar angle range from $8^{o}$ to $172^{o}$, which is the angular acceptance of the central tracker (TPC) at CEPC \cite{CEPC_CDR}. Events are generated simulating individually effects like the Initial State Radiation (ISR) and detector angular resolution, in order to study their impact on the $s'$ distribution with respect to the concurrent beam energy spread. Detector energy resolution is simulated by performing Gaussian smearing of the muons' polar angles w.r.t. the 0.1 mrad RMS corresponding to 100 $\mu$m position resolution foreseen for TPC at CEPC.\\

\begin{figure}[tbp]
	\centering 
	\includegraphics[width=1.\textwidth]{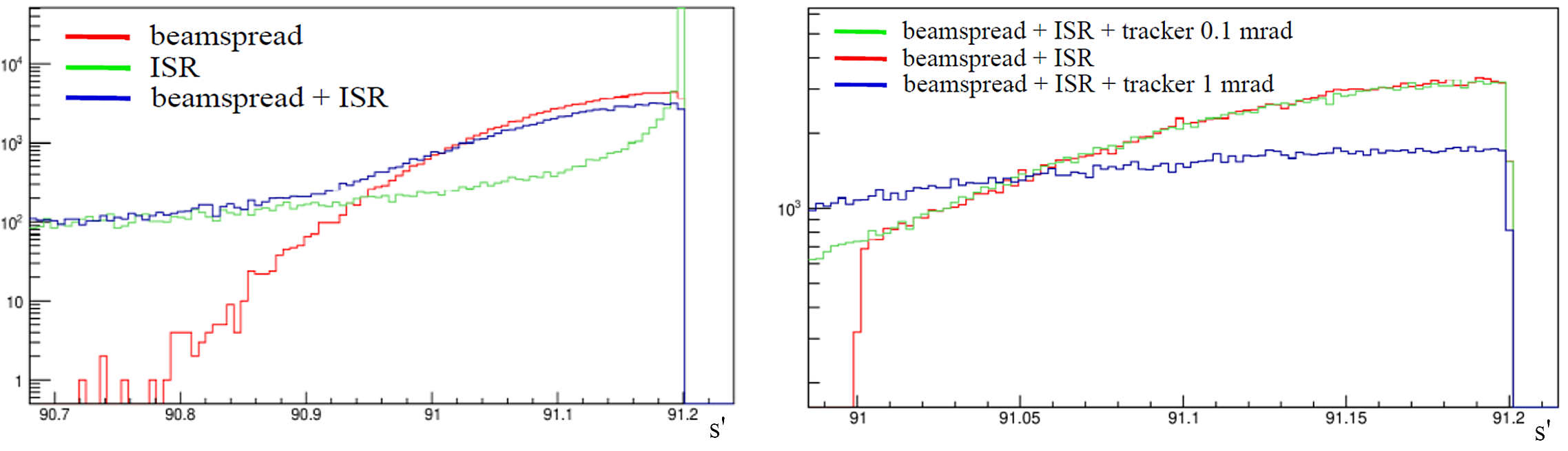}
	\hfill
	\caption{\label{fig:7} Left: Beam energy spread is the dominant effect to reduce the number of events at the maximal CM energy; Right: \textit{s'} sensitivity to the beam energy spread and different tracker resolutions.}
\end{figure}

The effective center-of-mass energy, $s'$, can be calculated from the reconstructed muons' polar angles \cite{sprime}, as:

\begin{equation}
\label{eq:2}
\frac{s'}{s}=\frac{\sin\theta^{+}+\sin\theta^{-}-\lvert\sin(\theta^{+}+\theta^{-})\rvert}{\sin\theta^{+}+\sin\theta^{-}+\lvert\sin(\theta^{+}+\theta^{-})\rvert}
\end{equation}

relying on the excellent TPC spatial resolution. As illustrated in Figure \ref{fig:7} (left), beam energy spread dominates the $s'$ shape at energies close to the nominal center-of-mass energy, while the Figure \ref{fig:7} (right) illustrates the effect of muon polar angle resolutions of 0.1 mrad and 1 mrad on top of ISR and the beam energy spread. From Figure \ref{fig:7} (right) is clear that tracker resolution of 0.1 mrad does not affect the $s'$ sensitivity to the beam energy spread. On the other hand, tracker resolution of 1 mrad significantly influences the method. We found that central tracker resolution in polar angle should not be larger than 0.5 mrad, corresponding to the 500 $\mu$m position resolution of TPC. The same stands for the beam energy of 120 GeV, where the nominal beam energy spread of 0.134\% is foreseen \cite{CEPC_CDR}. \\

\begin{figure}[tbp]
	\centering 
	\includegraphics[width=.45\textwidth]{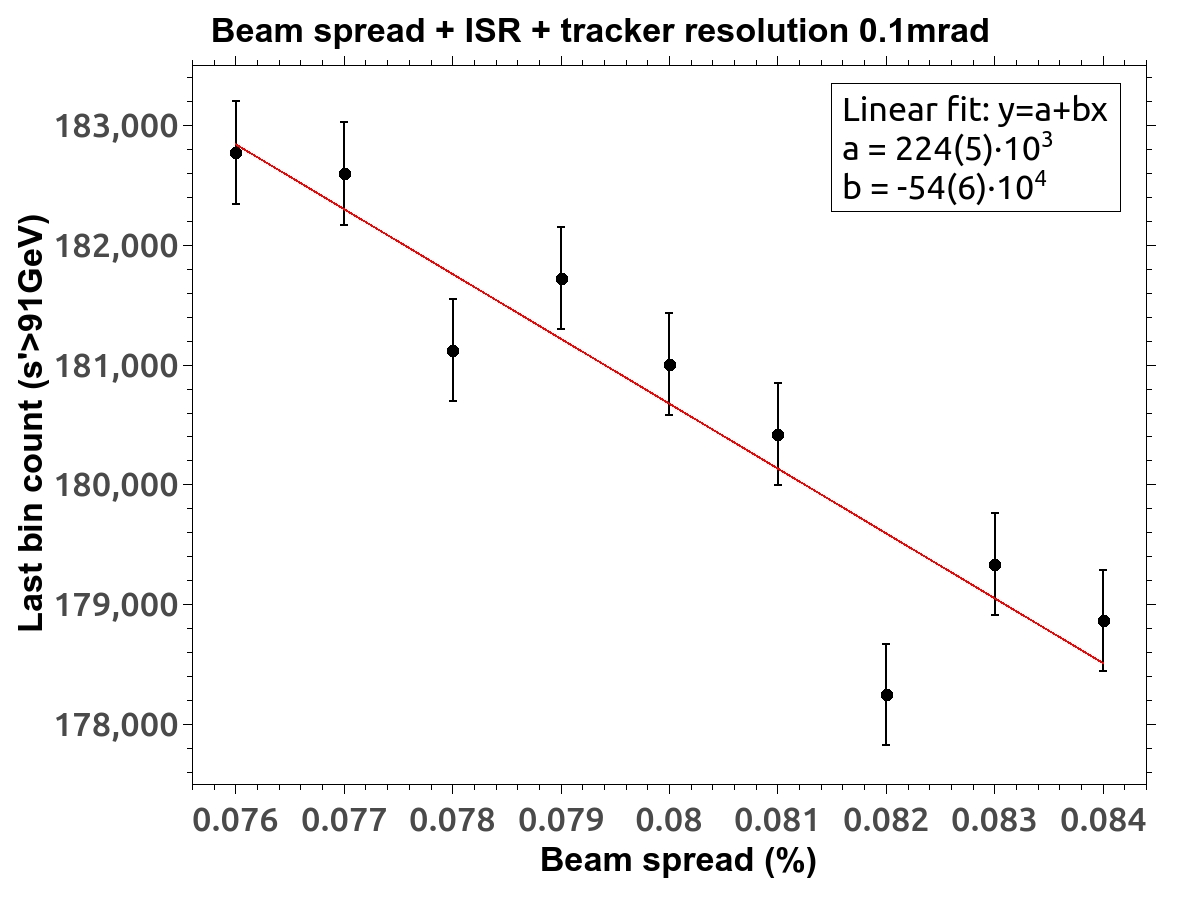}
	\hfill
	\includegraphics[width=.45\textwidth]{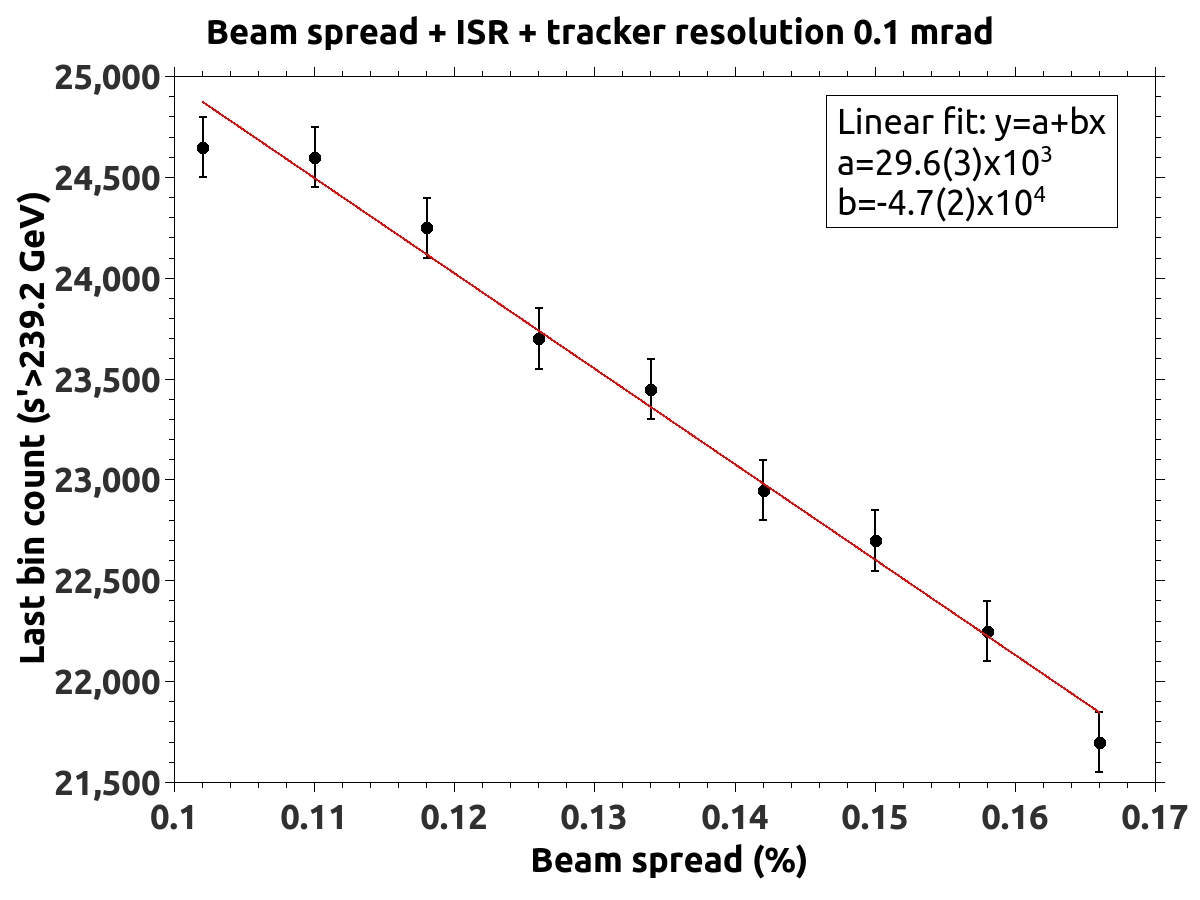}
	\caption{\label{fig:8} Left: Number of di-muon events at the maximal CM energy around nominal 91.2 GeV w.r.t. the beam energy spread; Right: The same at 240 GeV.}
\end{figure}

To exploit $s'$ peak sensitivity to the beam energy spread values, beam energy spread is varied around the nominal value, generating each time $2.5\cdot10^{5}$ ($10^{5}$) events per beam energy spread variation at 91.2 GeV (240 GeV). Dependence is illustrated in Figure \ref{fig:8}, at the $Z^{0}$ pole and 240 GeV. As expected, larger beam energy spread leads to the corresponding reduction of the number of di-muon events carrying close-to-maximal available energy from the collision. Knowing this dependence from simulation enables the determination of the effective beam energy spread ($\delta'$) once the count of di-muon events is known experimentally. \\

Table \ref{tab:BeamSpread} shows that with the assumed statistics of collected events at $Z^{0}$ pole, relative variations of the $\delta'$ can be measured with 25\% relative systematic uncertainty from the calibration curve and 1.2\% relative statistical uncertainty for only 3 minutes of data taking with $1.02 \cdot 10^{36}$ cm$^{-2}$s$^{-1}$ instantaneous luminosity. At 240 GeV center-of-mass energy, $\delta'$ can be measured with 15\% relative systematic uncertainty from the calibration curve and 2.3\% relative statistical uncertainty but for approximately 5 days of data taking with instantaneous luminosity of $5.2 \cdot 10^{34}$ cm$^{-2}$s$^{-1}$.\\

\begin{table}[!htp]
	\centering
	\caption{Beam energy spread variations experimentally accessible at CEPC.}
	\footnotesize
	\label{tab:BeamSpread}
	\begin{tabular}{|c|c|c|c|c|c|c|c|}
		\hline 
		\textbf{CEPC} & \vtop{\hbox{\strut \textbf{$\mathcal{L}$ @ IP}} \hbox{\strut \textbf{(cm$^{-2}$s$^{-1}$)}}} & \vtop{\hbox{\strut \textbf{Nominal}} \hbox{\strut \textbf{beam}} \hbox{\strut \textbf{energy}} \hbox{\strut \textbf{spread}} \hbox{\strut $\delta$ (\%)}} & \vtop{\hbox{\strut \textbf{Number}} \hbox{\strut \textbf{of}} \hbox{\strut \textbf{events}}} & \vtop{\hbox{\strut \textbf{Cross-}} \hbox{\strut \textbf{section}} \hbox{\strut\textbf{$e^{+}e^{-}\rightarrow \mu^{+}\mu^{-}$}}} & \vtop{\hbox{\strut \textbf{Collect.}} \hbox{\strut \textbf{time}}} &  \vtop{\hbox{\strut \textbf{Total}} \hbox{\strut \textbf{relative}} \hbox{\strut \textbf{uncert.}} \hbox{\strut \textbf{of $\delta'$}}} & \vtop{\hbox{\strut \textbf{Total}} \hbox{\strut \textbf{absolute}} \hbox{\strut \textbf{unc. of}} \hbox{\strut \textbf{$\delta'$ (MeV)}}} \\ 
		\hline 
		\textbf{$Z^{0}$ pole} & $1.02 \cdot 10^{36}$ & 0.080 & $2.5 \cdot 10^{5}$ & 1.5 nb & 3 min & 25\% & 9 \\ 
		\hline 
		\textbf{240 GeV} & $5.2 \cdot 10^{34}$ & 0.134 & $1.0 \cdot 10^{5}$ & 4.1 pb & 5 days & 15\% & 24 \\		
	\hline
\end{tabular} 
\end{table}

\begin{figure}[tbp]
	\centering 
	\includegraphics[width=0.45\textwidth]{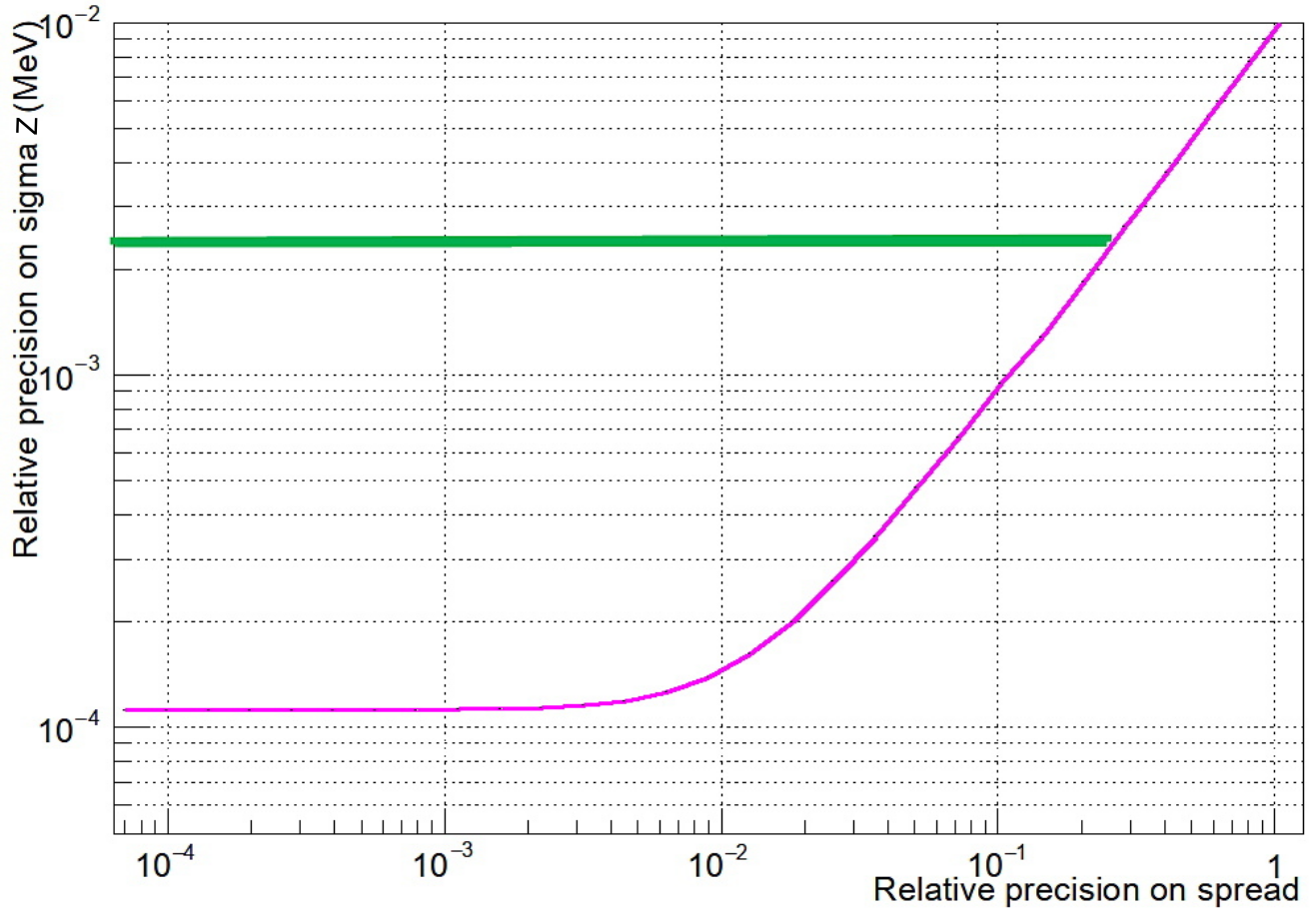}
	\hfill
	\caption{\label{fig:9} Impact of the relative precision of the beam energy spread on the $Z^{0}$ production cross-section.}
\end{figure}

\begin{figure}[tbp]
	\centering 
	\includegraphics[width=.45\textwidth]{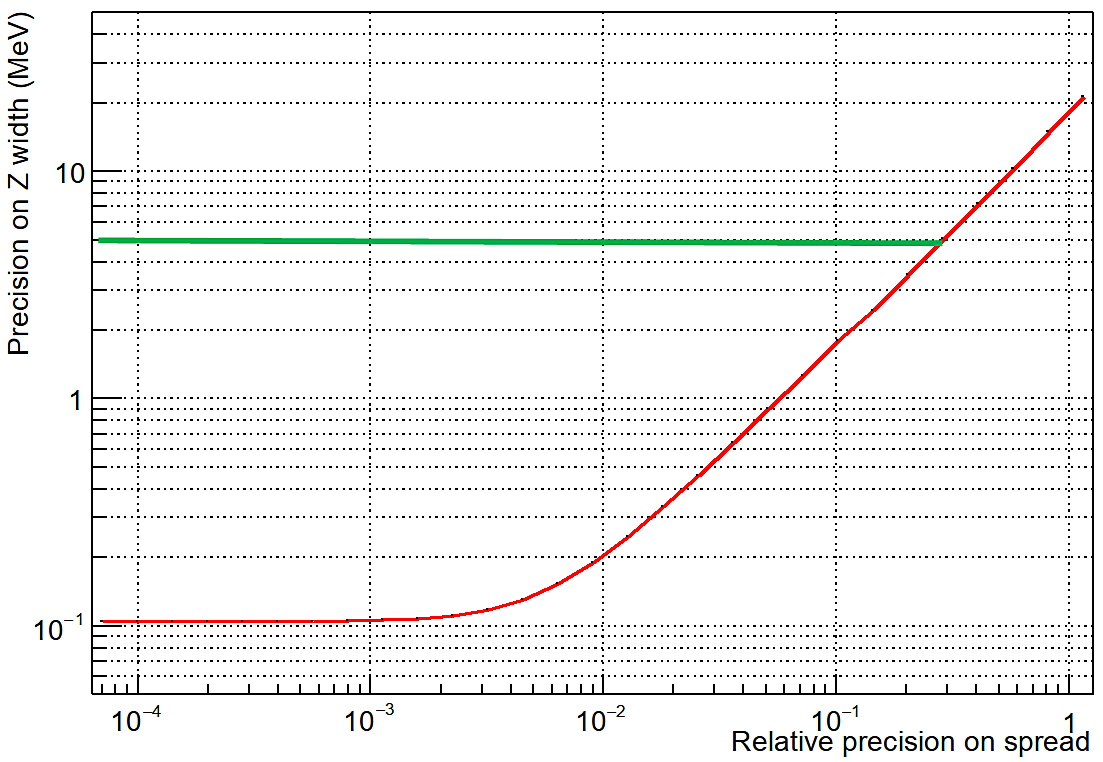}
	\hfill
	\includegraphics[width=.45\textwidth]{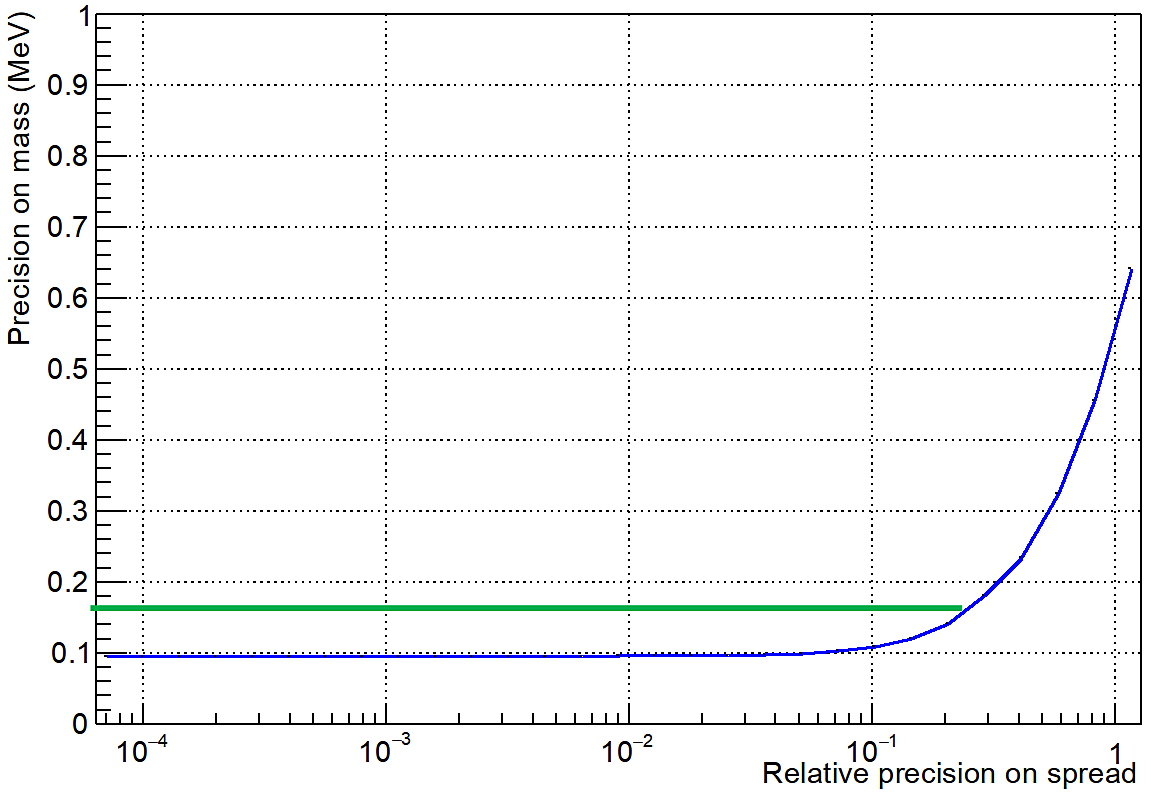}
	\caption{\label{fig:10} Impact of the relative precision of the beam energy spread on the $Z^{0}$ total width and mass absolute precisions.}
\end{figure}

As already mentioned, several precision EW observables at the $Z^{0}$ pole depend on the beam energy spread uncertainty. Figures \ref{fig:9} and \ref{fig:10} illustrate that the cross-section for $Z^{0}$ production ($\sigma_{Z}$), $Z^{0}$ total width ($\Gamma_{Z}$) and mass ($m_{Z}$) will receive following contributions from the foreseen beam energy spread uncertainty: $\delta(\sigma_{Z}) \sim 3 \cdot 10^{-3}$, $\Delta \Gamma_{Z} \sim$5 MeV and $\Delta m_{Z} \sim$160 keV.\\

Several refinements of this study are possible, yet not expected to significantly modify the results: effect of ISR (theoretical) uncertainty, FSR of muons, full detector simulation and impact of similar final state backgrounds, as well as presence of the beamstrahlung that is however much less pronounced at circular than at linear machines. \\

\section{Conclusions}
\label{sec:conclusions}

The study confirmed that the uncertainty of the luminometer inner radius at the micron level together with the uncertainties of the available center-of-mass energy and beam energy asymmetry below 10 MeV are posing the most challenging requirements on measurement of the integrated luminosity with $10^{-4}$ relative precision at the $Z^{0}$ pole. Permille precision of the integrated luminosity measurement at 240 GeV CEPC seems to be feasible from the point of view of the existing technologies. \\

It is also shown that with the CEPC post CDR design, beam energy spread can be determined with absolute accuracy corresponding to 9 MeV, in only 3 minutes of data-taking of $e^{+}e^{-} \rightarrow \mu^{+}\mu^{-}$ events at the $Z^{0}$ pole, where the statistical uncertainty contributes at the level of several hundreds of keV. The above translates to relative uncertainty of the $Z^{0}$ production cross-section of $3 \cdot 10^{-3}$ and absolute precisions of the $Z^{0}$ mass of 160 keV and width of 5 MeV respectively. \\

\acknowledgments
This research is supported by the Ministry of Education, Science and Technological Development
and by the Science Fund of the Republic of Serbia, through Grant No 7699827, HIGHTONE-P.
This paper also contains results accomplished in in collaboration of the authors with S. Lukic, previously in the Vinca Institute HEP Group.

\end{document}